\documentclass[]{ceurart}

\usepackage{url}  
\usepackage{listings}
\usepackage{tcolorbox}
\usepackage{multirow} 
\usepackage{parskip}
\usepackage{tabularx}
\usepackage{caption}
\usepackage{graphicx}
\usepackage{subcaption}
\usepackage{array}
\usepackage{amsmath,amssymb,amsfonts}
\usepackage{algorithmic}
\usepackage{textcomp}
\usepackage{xcolor}

\copyrightyear{2025}
\copyrightclause{Copyright © 2025 for this paper by its authors. Use permitted under Creative Commons License Attribution 4.0 International (CC BY 4.0).}
\begin{document}

\conference{}

\title{From Inductive to Deductive: LLMs-Based Qualitative Data Analysis in Requirements Engineering}

\author[1]{Syed Tauhid Ullah Shah}[%
email=syed.tauhidullahshah@ucalgary.ca,
]

\address[1]{ University of Calgary, Calgary, Canada}

\author[1]{Mohamad Hussein}[
email=mohamad.hussein@ucalgary.ca,
]

\author[1]{Ann Barcomb}[
email=ann.barcomb@ucalgary.ca,
]
\fnmark[1]

\author[2]{Mohammad Moshirpour}[
email=mmoshirp@uci.edu,
]

\address[2]{University of California, Irvine, Irvine, California, USA,}

\begin{abstract}
  Requirements Engineering (RE) is essential for developing complex and regulated software projects. Given the challenges in transforming stakeholder inputs into consistent software designs, Qualitative Data Analysis (QDA) provides a systematic approach to handling free-form data. However, traditional QDA methods are time-consuming and heavily reliant on manual effort. In this paper, we explore the use of Large Language Models (LLMs), including GPT-4, Mistral, and LLaMA-2, to improve QDA tasks in RE. Our study evaluates LLMs' performance in inductive (zero-shot) and deductive (one-shot, few-shot) annotation tasks, revealing that GPT-4 achieves substantial agreement with human analysts in deductive settings, with Cohen’s Kappa scores exceeding 0.7, while zero-shot performance remains limited. Detailed, context-rich prompts significantly improve annotation accuracy and consistency, particularly in deductive scenarios, and GPT-4 demonstrates high reliability across repeated runs. These findings highlight the potential of LLMs to support QDA in RE by reducing manual effort while maintaining annotation quality. The structured labels automatically provide traceability of requirements and can be directly utilized as classes in domain models, facilitating systematic software design."
\end{abstract}

\begin{keywords}
Requirements Engineering \sep Qualitative Data Analysis \sep Large Language Models \sep Zero-shot Learning \sep Few-shot Learning \sep Natural Language Processing
\end{keywords}

\newcommand{\ab}[1]{{\color{red}{\textbf{AB}: #1}}}

\maketitle

\section{Introduction}

Requirements Engineering is a key process in developing large and complex software systems. It ensures that the software meets the needs of stakeholders by gathering, organizing, and managing their requirements systematically \cite{cheng2007research}. QDA is an emerging approach in RE that aids in analyzing unstructured data like interviews and surveys to identify patterns and insights \cite{carrizo2014systematizing, mucha2023qdacity,kaufmann:2021:validation}. One important step in QDA is labeling or coding, where pieces of text are categorized into themes to make the data more structured and meaningful \cite{saldana2021coding}. This process helps improve traceability, consistency, and the quality of software design \cite{treude2024qualitative}. However, traditional QDA methods can be slow, inconsistent, and require a lot of manual work \cite{tsang2020experiment}. 

Recently, Large Language Models (LLMs), such as GPT-4 \cite{achiam2023gpt}, Gemini \cite{team2023gemini}, and LLaMA-2 \cite{touvron2023llama}, have shown great potential at processing and generating human-like text, making them useful for working with large sets of unstructured data. Unlike traditional models, LLMs use natural language prompts for tasks such as text classification \cite{sun2023text}, summarization \cite{zhang2024benchmarking}, and translation \cite{zhang2023prompting}. Their adaptability across zero-shot and few-shot scenarios \cite{brown2020language,team2023gemini} reduces reliance on extensive training data and computational resources. In RE, structured outputs like software specifications are essential, and LLMs can help by generating accurate and contextually relevant outputs \cite{krishna2024using}.

In this study, we use LLMs, such as GPT-4, Mistral, and LLaMA-2, to assist in qualitative data annotation for RE, aiming to reduce manual effort and accelerate the analysis process. Our approach uses both inductive and deductive annotation. To facilitate the alignment of inductive and deductive with the NLP setup, we treated inductive annotation as zero-shot learning and used one-shot and few-shot learning for deductive annotation. Our experiments, conducted on two test cases (Library Management System and Smart Home System), demonstrate that our LLM-based approach achieved fair to substantial agreement with human analysts in deductive annotation tasks. Specifically, in both test cases, GPT-4 performed better than the other LLMs, showing stronger agreement with human analysts. Contextual examples in detailed prompts led to notable performance gains, especially during the shift from zero-shot to one-shot scenarios. Providing rich context was key, as it produced much better results than using limited or no context. Our findings demonstrate that LLMs can effectively support qualitative data annotation in RE, offering faster and more consistent results. Additionally, the structured labels generated by these models help create domain models, which are critical for systematic software design and development. This not only reduces manual effort but also ensures greater consistency and accuracy, improving the overall quality of software design.

Our work is structured around the following research key questions:

\begin{itemize}
    \item \textbf{RQ1:} To what extent does our LLM-based approach align with human analysts in both inductive and deductive annotation tasks?
    \item \textbf{RQ2:} How do different prompt designs (zero-shot and few-shot) and lengths (short, medium, long) affect the accuracy and reliability of the annotations generated by LLMs?
    \item \textbf{RQ3:} How consistent are the LLM-generated labels across multiple runs?
    \item \textbf{RQ4:} How do various contextual settings affect the effectiveness of our LLM-based annotation approach?
\end{itemize}

Overall, our contributions can be summarized as follows:
\begin{itemize}
    \item We conducted a comprehensive assessment of both open-source and proprietary LLMs to determine their utility in supporting QDA within RE. Our study spans various models, including GPT-4, Mistral, and LLaMA-2.
    
   \item We explored the effectiveness of different annotation strategies (inductive and deductive) across various settings (zero-shot, one-shot, and few-shot). Our findings illustrate the impacts of these strategies on the performance of LLMs, with deductive (few-shot) annotation achieving higher agreement with human analysts. For instance, GPT-4 reached a Cohen’s Kappa score of up to 0.738, indicating substantial agreement.

    \item  We investigated the influence of prompt length and contextual information on the performance of LLMs. Detailed, context-rich prompts significantly enhanced the accuracy of LLMs. In the few-shot setting, the precision and recall for GPT-4 were notably high, at 0.80 and 0.79, respectively, demonstrating its effectiveness in closely mirroring human analytical processes.

\end{itemize}

\section{Literature Review}

In this literature review, we explore two critical areas: the role of QDA in RE (Section. \ref{lit:qda-re}) and the application of LLMs in RE (Section. \ref{lit:llm-re}) for QDA-assisted RE.

\subsection{Qualitative Data Analysis (QDA)-based RE}
\label{lit:qda-re}

QDA is a key technique in RE for analyzing unstructured stakeholder inputs, such as interviews and surveys, to extract patterns and generate actionable insights \cite{nuseibeh2000requirements}. Qualitative labeling is used to identify domain concepts and latent requirements. These coded insights are then mapped to classes or components in a domain model, ensuring that stakeholder needs are accurately reflected in the system design \cite{kaufmann2022validation}.  While QDA improves traceability and accuracy in requirements specification, traditional methods are labor-intensive, inconsistent, and prone to subjectivity \cite{chen2016challenges, glaser2017discovery}. Tools like Computer Assisted Qualitative Data Analysis Software (CAQDAS) aim to support the process but often lack adaptability to dynamic RE environments \cite{kaufmann2019qdacity}. Recent efforts like QDAcity-RE \cite{kaufmann2019qdacity, kaufmann:2020:supporting} have shown that QDA techniques help extract domain concepts from unstructured stakeholder interviews and documentation. This approach uses manual qualitative coding to generate traceable domain models by mapping labeled requirements to classes or components, ensuring consistency and traceability in the design process. However, the repetitive and manual nature of these processes underscores the need for automation to improve scalability and efficiency.

\subsection{Large Language Models (LLMs) in Requirements Engineering (RE)}
\label{lit:llm-re}
 LLMs, such as GPT-4, Mistral, and LLaMA-2, have shown promise in automating RE tasks like requirements classification, ambiguity detection, and documentation synthesis \cite{vogelsang2024using, fan2023large}. Their adaptability across zero-shot and few-shot scenarios enables efficient processing of unstructured data with minimal training \cite{brown2020language}. Recent studies have explored the application of LLMs in qualitative research within software engineering \cite{bano2024large}. For example, \citeauthor{alhoshan2023zero} \cite{alhoshan2023zero} demonstrated the potential of LLMs for requirements classification without task-specific training, while \citeauthor{kici2021bert} \cite{kici2021bert} showed the effectiveness of transfer learning for RE tasks. Despite this progress, applying LLMs to QDA for RE remains underexplored, presenting an opportunity to address limitations of traditional QDA and enhance scalability and accuracy in RE processes.

Although LLMs have been widely studied in RE and QDA independently, their integration for QDA in RE is still new. Using LLMs for QDA can greatly improve efficiency and accuracy by automating annotations and reducing errors from manual work, can simplify the process, make it more reliable and scalable, and better meet the changing demands of RE.

\section{Qualitative Data Analysis (QDA)}
\label{sec:QDA}

For our study, we focused on two specific test cases: a Library Management System and a Smart Home system. The Library Management System test case involves managing resources like cataloging, user management, loans, and digital resources. The Smart Home System test case focuses on automating tasks such as security, energy control, and device management. While the two primary test cases were sourced from the PURE dataset \cite{pure}, we supplemented these with additional SRS and FRS documents from the internet to ensure a comprehensive dataset. Following the extensive data collection, we applied QDA to our test cases. Our primary goal was to convert the requirement statements from these documents into actionable insights by assigning precise labels to distinct segments. These labels, akin to UML classes, help structure the requirements, making them more comprehensible and aiding their integration into the software development lifecycle. This structured approach ensures that the requirements are clear, precise, and aligned with the overall goals of the software engineering process.

To maintain precision and reliability, we assigned two independent analysts, $C_1$ and $C_2$, to review and label the same set of requirement documents independently. Both analysts have a software engineering background, with $C_1$ having 1.5 years of experience and $C_2$ having 8 months of experience working with software requirements. First, both analysts ($C_1$ and $C_2$)  labeled the requirement documents independently. We then measured their agreement using Cohen’s Kappa \footnote{Cohen's Kappa is a statistical measure used to assess the inter-rater agreement of qualitative (categorical) items. It considers the agreement occurring by chance and provides a more robust metric compared to simple percent agreement. A Kappa score of more than 0.70 typically indicates a substantial level of agreement between raters, reflecting a high degree of reliability in the annotation process.}. After that, they met to discuss and resolve any differences, creating a unified set of labels. This iterative process combined their insights into a unified analytical framework. The total time and effort spent by the analysts in this QDA process are summarized in Table. \ref{tab:time_spent}. We reached a substantial agreement of 0.80 for the Library Management System and 0.78 for the Smart Home System. The Library Management System used labels such as 'Notification,' 'Loan,' 'Reservation,' 'Catalog,' etc., while the Smart Home System included 'Sensor,' 'Light,' 'Thermostat,' 'Device,' etc. These labels ensure stakeholder inputs are directly linked to corresponding elements in the domain model

\begin{table}[h]
\renewcommand{\arraystretch}{1.0}
    \centering
    \caption{Time spent for the RE-QDA annotation process}
    
    \label{tab:time_spent}
    \begin{tabular}{ll}
        \hline
        \textbf{Parameter} & \textbf{Amount/Duration} \\
        \hline
        Analyst $C_1$ & 37 hours \\  
        Analyst $C_2$ & 29 hours \\  
        Meeting Duration & 8 hours (12 meetings in total) \\  
        Entire Time Spent & 74 hours \\
        \hline
    \end{tabular}
\end{table}

\section{Methodology}

\subsection{Overview}

\begin{figure}[htb]
\centering
\vspace{-0.3cm}
\captionsetup{aboveskip=0pt, belowskip=0pt} 
\includegraphics[height=0.12\textheight, trim=0 0 0 0, clip]{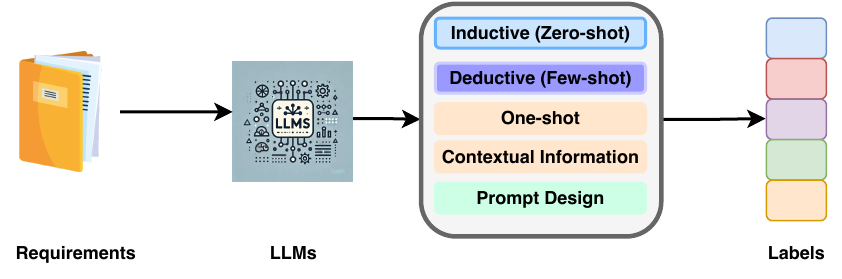}
\caption{Overview of the methodology integrating LLMs into QDA for RE. The process includes collecting requirement statements, designing prompts, feeding them to LLMs, and generating output labels.}
\label{fig:methodology}
\end{figure}

Figure. \ref{fig:methodology} outlines our approach to integrating LLMs into QDA for RE. We begin by taking requirement statements (Section. \ref{sec:QDA}) as input. The requirements are subsequently formatted into structured prompts optimized for inductive or deductive annotations (Section. \ref{subsec:promptdesign}). Inductive prompts, used in zero-shot learning, allow LLMs to identify patterns without predefined categories, while deductive prompts, supporting one-shot and few-shot learning, include examples for consistency with defined categories. LLMs (Section. \ref{subsec:model}) process these prompts to generate structured labels (Section. \ref{subsec:output}), which categorize and interpret requirements, providing actionable insights for further development. This approach simplifies the QDA process, reducing manual effort while leveraging LLM capabilities effectively. 

\vspace{-0.4cm}
\subsection{Prompt Design}
\label{subsec:promptdesign}

We created clear and structured prompts to convert the collected requirements into a format that LLMs can understand and label. Table~\ref{tab:prompts} summarizes our prompt templates, while Table~\ref{tab:context_info} provides details on our context levels. Our design considers three independent factors:

\textbf{1. Shot Type:}  
This factor refers to the number of examples included in the prompt. In a \textit{zero-shot} prompt, no examples are provided, so the LLM relies entirely on its built-in knowledge. A \textit{one-shot} prompt includes one example to guide the model, while a \textit{few-shot} prompt provides several examples to clearly show the desired labeling.

\begin{table}[htb]
\centering
\caption{Summary of Prompt Templates by Shot Type, Length, and Contextuality}
\label{tab:prompts}
\resizebox{0.95\textwidth}{!}{ 
\begin{tabular}{l l p{12cm}} 
\toprule
\textbf{Category} & \textbf{Prompt Type} & \textbf{Prompt Description (Example Template)} \\
\midrule
\multirow{3}{*}{\textit{Zero-shot Prompts (Inductive)}} 
    & \textbf{Short} & "Analyze this \{requirement\} and respond with ONLY a single Qualitative Data Analysis-based label." \\ \cline{2-3}
    & \textbf{Medium} & "You are analyzing software requirements for the \{system\_type\} system. Respond with ONLY a single category label that best captures the main functionality for the following \{requirement\}." \\ \cline{2-3}
    & \textbf{Long} & "You are performing Qualitative Data Analysis on requirements for a \{system\_type\} system. \newline \{context\}. Analyze the requirement below and respond with ONLY a single categorical label (1 word) that best represents its main functionality for the following \{requirement\}." \\ 
\midrule
\multirow{3}{*}{\textit{Few-shot Prompts (Deductive)}} 
    & \textbf{Short} & "Analyze requirements and respond with ONLY a single Qualitative Data Analysis-based label. \newline Examples: \{example1\}, \{example2\}, \{example3\}." \\ \cline{2-3}
    & \textbf{Medium} & "For a \{system\_type\} system, respond with ONLY a single Qualitative Data Analysis-based label that best represents the functionality. \newline Examples: \{example1\}, \{example2\}, \{example3\}." \\ \cline{2-3}
    & \textbf{Long} & "You are performing Qualitative Data Analysis on requirements for a \{system\_type\} system. \newline \{context\} \newline Given the following examples: \newline Example 1: \{example1\} (Label: \{label1\}) \newline Example 2: \{example2\} (Label: \{label2\}) \newline Example 3: \{example3\} (Label: \{label3\}) \newline Analyze the requirement below and respond with ONLY a single Qualitative Data Analysis-based label (1 word) that represents its main functionality." \\ 
\bottomrule
\end{tabular}
} 
\end{table}

\textbf{2. Prompt Length:}  
This factor measures how much instruction is given. A \textit{short} prompt provides minimal instructions, a \textit{medium} prompt adds additional details, and a \textit{long} prompt gives in-depth guidance. For instance, a long prompt might explain specific aspects of QDA such as traceability, stakeholder intent, and consistency.

\textbf{3. Contextual vs. Non-Contextual:}  
This aspect determines whether the prompt includes background information. Non-contextual prompts provide only the requirement statement, while contextual prompts offer system details to improve understanding. We define three levels: no context (requirement only), some context (brief system description), and full context (comprehensive system details).

\subsection{Model Selection}
\label{subsec:model}
\begin{table}[htb]
\centering
\caption{Context Levels for Prompt Design}
\label{tab:context_info}
\begin{tabular}{lp{10cm}} 
\toprule
\textbf{Context Level} & \textbf{Description and Example} \\
\midrule
\textbf{No Context} & Only the requirement is provided. \\ 
                    & Example: "Requirement: \{requirement\}" \\
\midrule
\textbf{Some Context} & A brief system description is added. \\ 
                      & Example: "This is a Library Management System that handles cataloging, user management, and loans. Requirement: \{requirement\}" \\
\midrule
\textbf{Full Context} & A comprehensive system description is provided, detailing functionalities and design specifics. \\ 
                      & Example: "The Library Management System (LMS) manages all aspects of a modern library, including resource cataloging, loan processing, digital resource management, and administrative reporting. Requirement: \{requirement\}" \\
\bottomrule
\end{tabular}
\end{table}

We used state-of-the-art LLMs, including GPT-4 \cite{achiam2023gpt}, Mistral \cite{jiang2023mistral}, and LLaMA-2 \cite{touvron2023llama}, for their abilities in understanding and generating natural language and suitability for the complex task of QDA in RE. We prompt these models with specific software requirement data to understand the context of requirements, recognize domain-specific terminology, and map requirement statements to relevant labels. 

\subsection{Output Labels}
\label{subsec:output}

Our approach focuses on generating labels that organize and interpret requirement statements, converting unstructured data into clear and actionable insights. These labels are critical for understanding stakeholder needs and ensuring that requirements align with their expectations \cite{wiegers2013software}. By improving communication among teams, the labels also play a key role in creating domain models, which are essential for systematic software design \cite{kaufmann:2020:supporting}. To achieve accurate and relevant labels, we employ both inductive and deductive strategies, supported by contextual prompts.  This dual strategy improves the precision and relevance of the labeling process. Additionally, these QDA-based annotations ensure automatic traceability by linking each label back to its corresponding stakeholder input \cite{kaufmann2019qdacity}.

\section{Results}

\subsection{Evaluation Metrics}
We assessed the performance of the LLMs using several key metrics to evaluate their accuracy and agreement in annotation tasks. Inter-rater agreement was measured using Cohen's Kappa, which quantifies the level of agreement between the labels generated by the LLMs and those assigned by human analysts, with higher values indicating stronger agreement. To evaluate the consistency of the labels across multiple experimental runs, we analyzed the standard deviation (SD) and the Intraclass Correlation Coefficient (ICC). A lower SD indicates minimal variability in the labels, while ICC values above 0.85 demonstrate excellent reliability. In addition to reliability and consistency, we evaluated the accuracy of the LLMs, which measures the proportion of correct labels out of all predictions. Precision was used to determine how many of the labels identified by the LLMs were correct, providing insights into their ability to avoid false positives. Recall assessed the ability of LLMs in the identification of all relevant labels, minimizing the risk of missing important instances (false negatives). Finally, we used F1-score, the harmonic mean of precision and recall, to provide a balanced measure of the performance of the models, with higher scores indicating a good trade-off between precision and recall. In this study, we used only the labels on which both analysts reached consensus as the ground truth for evaluating LLM performance.

\begin{table*}
  \caption{Comparison of Cohen's Kappa Scores for Different Models Across Both Test Cases}
  \label{tab:comp}
  \centering
  \begin{tabular}{llccc}
    \toprule
    \textbf{Test Case} & \textbf{Setting} & \textbf{Llama 2} & \textbf{Mistral} & \textbf{GPT-4} \\
    \midrule
    \multirow{3}{*}{\textbf{Library Management System}} 
      & Zero-shot & 0.516 & 0.526 & \textbf{0.543} \\
      & One-shot  & 0.675 & 0.685 & \textbf{0.690} \\
      & Few-shot  & 0.730 & 0.734 & \textbf{0.738} \\
    \midrule
    \multirow{3}{*}{\textbf{Smart Home System}} 
      & Zero-shot & 0.514 & 0.530 & \textbf{0.541} \\
      & One-shot  & 0.681 & 0.686 & \textbf{0.689} \\
      & Few-shot  & 0.728 & 0.730 & \textbf{0.734} \\
    \bottomrule
  \end{tabular}
\end{table*}

\subsection{Implementation}
We carried out all experiments using Python and PyTorch\footnote{\url{https://pytorch.org/}}. For Mistral and LLaMA-2 models, we used the 7B configuration from the Hugging Face's Transformers library\footnote{\url{https://huggingface.co/transformers/}}, which provides access to pre-trained models while for GPT-4, we used the GPT-4 Turbo API\footnote{\url{https://platform.openai.com/docs/models/gpt-4-turbo-and-gpt-4}}. To ensure fair comparisons, we set the temperature parameter to 0.0 across all models, which minimizes randomness and makes outputs consistent. The experiments were conducted on high-performance computing clusters equipped with NVIDIA A100 GPUs to handle the computational demands. The source code for all experiments and evaluations is publicly available\footnote{\url{https://github.com/SyedTauhidUllahShah/LLM4QDARE}}.

\subsection{LLMs vs. Human Analysts (RQ1)}
To evaluate the effectiveness of LLMs in aiding QDA-based annotation tasks within RE, we compared their performance against human analysts for both inductive and deductive settings. We used Cohen’s Kappa, a widely recognized statistical measure for assessing inter-rater agreement, to quantify agreement levels between LLM-generated labels and those derived by human analysts (described in detail in Section. \ref{sec:QDA}). This measure highlights the reliability and consistency of the LLM's performance in replicating human judgment, aligning with practices in qualitative research \cite{coleman2024intercoder} and LLM-assisted content analysis \cite{chew2023llm}.

Table. \ref{tab:comp} reports the Cohen's Kappa results for various prompt designs (zero-shot, one-shot, few-shot) and test cases (Library Management System and Smart Home System). Our empirical assessment across various settings for both test-cases yielded significant insights into the capabilities of LLMs. Notably, GPT-4 consistently outperformed other models such as LLaMA-2 and Mistral, achieving the highest Cohen’s Kappa scores. Specifically, in the few-shot setting, GPT-4 achieved scores of 0.738 and 0.734 for the Library Management System and the Smart Home System, respectively, indicating substantial agreement with human analysts and highlighting its robustness in these settings.

\begin{table*}
  \caption{Comparison of Cohen's Kappa Scores for Models with different Prompt Lengths (Short, Medium, Long)}
  \label{tab:promptlength}
  \centering
  \begin{tabular}{lccc|ccc}
    \toprule
    \textbf{Model} & \multicolumn{3}{c}{\textbf{Library Management}} & \multicolumn{3}{c}{\textbf{Smart Home}} \\
    \cmidrule(lr){2-4} \cmidrule(lr){5-7}
                   & \textbf{Short} & \textbf{Medium} & \textbf{Long} & \textbf{Short} & \textbf{Medium} & \textbf{Long} \\
    \midrule
    Llama 2        & 0.629          & 0.686           & 0.707         & 0.624          & 0.698           & 0.705 \\
    Mistral        & 0.645          & 0.699           & 0.713         & 0.633          & 0.685           & 0.712 \\
    GPT-4          & 0.641          & 0.691           & \textbf{0.738} & 0.631          & 0.681           & \textbf{0.734} \\
    \bottomrule
  \end{tabular}
\end{table*}
However, it is important to note that the agreement levels in the zero-shot setting were around 0.54, which is not typically considered a strong outcome. This observation suggests that while LLMs can approach the performance of human analysts in scenarios where some guidance (one-shot or few-shot) is provided, their effectiveness in fully autonomous, inductive annotation tasks (zero-shot) remains limited. This analysis highlights that, although LLMs show promise, particularly in deductive settings where they can match or even exceed human performance, they still require refinement for inductive tasks where no initial guidance is given. This detailed understanding addresses \textbf{RQ1}, indicating that while LLMs hold significant potential to support human efforts in RE annotation processes, their current application is more reliable in deductive annotation tasks than inductive ones.

\subsection{Influence of Prompt Design on Annotation Outcomes (\textbf{RQ2})}

To assess the impact of different prompt lengths, we executed a series of experiments across the two distinct test cases. The results, summarized in Table. \ref{tab:promptlength}, indicate that while long prompts generally provide the best performance, medium prompts also offer a good balance of context and efficiency. Short prompts, although less detail-intensive, often fall short in tasks requiring detailed contextual understanding.

This analysis directly addresses \textbf{RQ2}, demonstrating that careful prompt design is essential for maximizing the effectiveness of LLMs in annotation tasks within RE. Also, the finding is consistent with the broader literature \cite{turpin2024language, wei2022chain}, which emphasizes that the detailed contextual information in long prompts significantly enhances LLM performance by reducing ambiguity. Our findings highlight the potential for optimizing LLM performance in practical applications by tailoring prompts to balance context and efficiency. 

\subsection{Consistency Analysis of LLM-Generated Labels Across Multiple Runs (\textbf{RQ3})}

The consistency analysis of LLM-generated labels across multiple runs, as shown in Table. \ref{tab:consistency_analysis}, revealed that GPT-4 exhibited the highest consistency among the tested models. Specifically, GPT-4 achieved the lowest standard deviations of 0.034 for the Library Management System and 0.037 for the Smart Home System. Additionally, GPT-4 obtained the highest ICC values of 0.93 and 0.92 for the Library Management System and Smart Home System, respectively. These results indicate a high degree of reliability and stability in the generated labels, surpassing the performance of LLaMA-2 and Mistral, which also demonstrated good consistency but with slightly higher variability.

The high ICC values (>0.85) across all models affirm that LLM-generated labels are consistently reproducible within the same class, ensuring reliable outputs that closely align with the performance of human analysts. These findings show that GPT-4 is a reliable tool for helping with QDA in RE, making it easier to extract and organize insights from requirements data with less manual work.

\begin{table*}
  \caption{Consistency Analysis of LLM-Generated Labels Across Multiple Runs for Both Test Cases}
  \label{tab:consistency_analysis}
  \centering
  \begin{tabular}{clccc}
    \toprule
    \textbf{Test Case} & \textbf{Metric} & \textbf{Llama 2} & \textbf{Mistral} & \textbf{GPT-4} \\
    \midrule
    \multirow{2}{*}{\textbf{Library Management}} 
    & \textbf{SD}  & 0.057 & 0.048 & \textbf{0.034} \\
    & \textbf{ICC} & 0.87  & 0.89  & \textbf{0.93} \\
    \midrule
    \multirow{2}{*}{\textbf{Smart Home}} 
    & \textbf{SD}  & 0.062 & 0.051 & \textbf{0.037} \\
    & \textbf{ICC} & 0.85  & 0.88  & \textbf{0.92} \\
    \bottomrule
  \end{tabular}
\end{table*}

\subsection{Impact of Contextual Backgrounds (\textbf{RQ4})}

To address \textbf{RQ4}, we evaluated the impact of varying levels of contextual backgrounds on the effectiveness of LLM-generated labels. Specifically, we adjusted the amount of context provided in the prompts, ranging from no context to full context. The results, as shown in Table. \ref{tab:context}, demonstrated that the inclusion of richer contextual information in the prompts significantly improved the performance of all evaluated models, including LLaMA-2, Mistral, and GPT-4.

Specifically, GPT-4 exhibited the highest Cohen's Kappa scores across all scenarios, achieving scores of 0.738 for the Library Management System and 0.734 for the Smart Home System in the full context setting. These findings indicate that GPT-4 is particularly effective at leveraging detailed contextual information to generate accurate and consistent labels.

The improvement in performance with increased context suggests that providing comprehensive background information enables LLMs to better understand and interpret the requirements, resulting in more precise annotation. This highlights the importance of designing context-rich prompts to maximize the potential of LLMs for automating and refining QDA processes within RE. By incorporating detailed contextual information, LLMs can deliver outputs that accurately reflect the complexities of the requirements, thereby improving the   accuracy and reliability of the annotation process.

 \begin{table*}
  \caption{Cohen's Kappa Analysis of Contextual Information Across different Levels of Contextual Information}
  \label{tab:context}
  \centering
  \begin{tabular}{clccc}
    \toprule
    \textbf{Test Case} & \textbf{Context} & \textbf{Llama 2} & \textbf{Mistral} & \textbf{GPT-4} \\
    \midrule
    \multirow{3}{*}{\textbf{Library Management}} 
    & \textbf{No Context}   & 0.663 & 0.674 & \textbf{0.712} \\
    & \textbf{Some Context} & 0.682 & 0.689 & \textbf{0.718} \\
    & \textbf{Full Context} & 0.707 & 0.713 & \textbf{0.738} \\
    \midrule
    \multirow{3}{*}{\textbf{Smart Home}} 
    & \textbf{No Context}   & 0.673 & 0.682 & \textbf{0.713} \\
    & \textbf{Some Context} & 0.691 & 0.701 & \textbf{0.722} \\
    & \textbf{Full Context} & 0.705 & 0.712 & \textbf{0.734} \\
    \bottomrule
  \end{tabular}
\end{table*}

\begin{table*}
  \caption{Detailed Performance Metrics for Different Models in Zero-shot, One-shot, and Few-shot Settings Across Test Cases}
  \label{tab:performance_metrics}
  \centering
  \resizebox{\textwidth}{!}{%
  \begin{tabular}{clcccccccc}
    \toprule
    \textbf{Setting} & \textbf{Model} & \multicolumn{4}{c}{\textbf{Library Management}} & \multicolumn{4}{c}{\textbf{Smart Home}} \\
    \cmidrule(lr){3-6} \cmidrule(lr){7-10}
    & & \textbf{Accuracy} & \textbf{Precision} & \textbf{Recall} & \textbf{F1-Score} 
      & \textbf{Accuracy} & \textbf{Precision} & \textbf{Recall} & \textbf{F1-Score} \\
    \midrule
    \multirow{3}{*}{Zero-shot} 
    & Llama 2 & 0.68 & 0.65 & 0.64 & 0.645 & 0.67 & 0.64 & 0.63 & 0.635 \\
    & Mistral & 0.70 & 0.68 & 0.67 & 0.675 & 0.69 & 0.66 & 0.65 & 0.655 \\
    & GPT-4   & 0.72 & 0.70 & 0.69 & 0.695 & 0.71 & 0.68 & 0.67 & 0.675 \\
    \midrule
    \multirow{3}{*}{One-shot} 
    & Llama 2 & 0.78 & 0.72 & 0.71 & 0.715 & 0.77 & 0.71 & 0.70 & 0.705 \\
    & Mistral & 0.80 & 0.74 & 0.73 & 0.735 & 0.79 & 0.73 & 0.72 & 0.725 \\
    & GPT-4   & 0.82 & 0.76 & 0.75 & 0.755 & 0.81 & 0.75 & 0.74 & 0.745 \\
    \midrule
    \multirow{3}{*}{Few-shot} 
    & Llama 2 & 0.84 & 0.76 & 0.74 & 0.750 & 0.83 & 0.75 & 0.73 & 0.740 \\
    & Mistral & 0.85 & 0.78 & 0.76 & 0.770 & 0.84 & 0.77 & 0.75 & 0.760 \\
    & GPT-4   & \textbf{0.86} & \textbf{0.80} & \textbf{0.79} & \textbf{0.795} 
              & \textbf{0.85} & \textbf{0.79} & \textbf{0.78} & \textbf{0.785} \\
    \bottomrule
  \end{tabular}%
  }
\end{table*}

\subsection{Performance Evaluation with Detailed Metrics}

To further validate our results, we incorporated additional evaluation metrics: accuracy, precision, recall, and F1-score. The detailed performance evaluation, presented in Table. \ref{tab:performance_metrics}, shows that GPT-4 consistently outperforms LLaMA-2 and Mistral across all metrics. Specifically, GPT-4 achieves the highest accuracy, precision, and recall in both zero-shot and few-shot settings for the Library Management and Smart Home test cases. Although in the inductive scenario, the model is not provided with explicit examples, it still outputs a single label per requirement that is evaluated against the ground truth. In the deductive scenario, implemented as few-shot learning, the model is guided by explicit examples to generate labels. In both cases, the task is treated as a multi-class classification problem. For instance, in the few-shot setting for the Library Management test case, GPT-4 achieves an accuracy of 0.86, a precision of 0.80, recall of 0.79, and an F1-score of 0.79, demonstrating its superior ability to correctly and consistently categorize requirement statements.

Similarly, in the Smart Home test case, GPT-4 again leads with an accuracy of 0.85, a precision of 0.79, a recall of 0.78, and an F1-score of 0.785 in the few-shot setting. This analysis supports our earlier findings from Cohen's Kappa and ICC, showing that GPT-4 is reliable for automating QDA tasks in RE. The higher precision and recall suggest that GPT-4 not only identifies the correct labels more often but also misses fewer important instances, making the annotations more complete and accurate.

\section{Threats to Validity}
In this section, we discuss the potential threats to the validity of our study on the application of LLMs for QDA in RE.

\subsection{Internal Validity} 
One challenge in this study is the potential bias in pre-trained LLMs such as GPT-4, Mistral, and LLaMA-2. Since these models are trained on vast datasets, their outputs may reflect underlying biases that could skew the annotation results and fail to fully capture the nuances of RE. To minimize this risk, we carefully designed prompts with detailed context to guide the models toward more accurate and relevant annotations. Another concern is the consistency of human annotations. Different analysts may interpret and label the same requirement statements in slightly different ways, which could introduce inconsistencies in the dataset used for evaluation. To address this, we used an inter-rater reliability phase, where analysts reviewed their annotations together, resolving discrepancies to improve label consistency. Prompt design also plays a crucial role in the accuracy of LLM-generated annotations. Poorly structured or vague prompts can lead to unreliable results. To improve performance, we tested prompts with different lengths and levels of contextual information, refining them through an iterative process to ensure clarity and effectiveness.

\subsection{External Validity}
Our study evaluates LLM performance using two test cases, Library Management and Smart Home systems, which may not fully capture the diversity of software systems in practice. Results could vary when applied to different domains, particularly those with unique complexities or highly specialized requirements. The dataset, while sourced from multiple documents, may not represent the full range of real-world projects. A broader selection of requirement documents covering various industries and project types would strengthen the evaluation and improve the generalizability of our findings. Contextual information in prompts also plays a key role in guiding LLMs toward accurate annotations, but our prompts may not fully capture every detail of different RE contexts. Ensuring clarity and relevance across diverse scenarios remains a challenge. Further refinement, incorporating real-world feedback, is needed to enhance the applicability of this approach.

\section{Conclusion and Future Work}

This paper explored the application of LLMs, specifically LLM, Mistral, and LLaMA-2, to aid and enhance the annotation processes in RE. Our findings demonstrate that GPT-4, in particular, significantly reduces the manual effort required for annotation, achieving high levels of accuracy and consistency comparable to human analysts.  The performance of these models is notably improved with detailed, context-rich prompts, underscoring the importance of prompt design in leveraging LLM capabilities effectively. Our work highlights that while GPT-4 and other LLMs show promise in deductive annotation tasks (one-shot and few-shot settings), achieving substantial agreement with human analysts, their effectiveness in inductive annotation tasks (zero-shot) remains limited. This calls for further development and optimization of LLM strategies to enhance their performance across all types of annotation tasks. The potential for broader adoption of LLMs in RE is clear, suggesting that these models can aid QDA, increase efficiency, and reduce subjectivity. The structured labels generated by LLMs not only improve the efficiency and reliability of the QDA process but also facilitate the creation of domain models, simplifying the software design process and enhancing overall project efficiency. Future work should focus on extending these results to more diverse scenarios and further refining the training processes to address any inherent model biases. By doing so, the utility and reliability of LLMs in enhancing various aspects of software development processes can be significantly expanded.

\bibliography{bibo}

\begin{thebibliography}{33}
\expandafter\ifx\csname natexlab\endcsname\relax\def\natexlab#1{#1}\fi
\providecommand{\url}[1]{\texttt{#1}}
\providecommand{\href}[2]{#2}
\providecommand{\path}[1]{#1}
\providecommand{\DOIprefix}{doi:}
\providecommand{\ArXivprefix}{arXiv:}
\providecommand{\URLprefix}{URL: }
\providecommand{\Pubmedprefix}{pmid:}
\providecommand{\doi}[1]{\href{http://dx.doi.org/#1}{\path{#1}}}
\providecommand{\Pubmed}[1]{\href{pmid:#1}{\path{#1}}}
\providecommand{\bibinfo}[2]{#2}
\ifx\xfnm\relax \def\xfnm[#1]{\unskip,\space#1}\fi
\bibitem[{Cheng and Atlee(2007)}]{cheng2007research}
\bibinfo{author}{B.~H. Cheng}, \bibinfo{author}{J.~M. Atlee},
\newblock \bibinfo{title}{Research directions in requirements engineering},
\newblock \bibinfo{journal}{Future of software engineering (FOSE'07)}  (\bibinfo{year}{2007}) \bibinfo{pages}{285--303}.
\bibitem[{Carrizo et~al.(2014)Carrizo, Dieste, and Juristo}]{carrizo2014systematizing}
\bibinfo{author}{D.~Carrizo}, \bibinfo{author}{O.~Dieste}, \bibinfo{author}{N.~Juristo},
\newblock \bibinfo{title}{Systematizing requirements elicitation technique selection},
\newblock \bibinfo{journal}{Information and Software Technology} \bibinfo{volume}{56} (\bibinfo{year}{2014}) \bibinfo{pages}{644--669}.
\bibitem[{Mucha(2023)}]{mucha2023qdacity}
\bibinfo{author}{J.~Mucha}, \bibinfo{title}{The QDAcity-RE-RS Method for Creating Complete, Consistent, and Traceable Requirements Specifications}, \bibinfo{publisher}{Friedrich-Alexander-Universitaet Erlangen-Nuernberg (Germany)}, \bibinfo{year}{2023}.
\bibitem[{Kaufmann et~al.(2021)Kaufmann, Krause, Harutyunyan, Barcomb, and Riehle}]{kaufmann:2021:validation}
\bibinfo{author}{A.~Kaufmann}, \bibinfo{author}{J.~Krause}, \bibinfo{author}{N.~Harutyunyan}, \bibinfo{author}{A.~Barcomb}, \bibinfo{author}{D.~Riehle},
\newblock \bibinfo{title}{A validation of {QDA}city--{RE} for domain modeling using qualitative data analysis},
\newblock \bibinfo{journal}{Requirements Engineering}  (\bibinfo{year}{2021}). \URLprefix \url{https://link.springer.com/article/10.1007/s00766-021-00360-6}. \DOIprefix\doi{https://doi.org/10.1007/s00766-021-00360-6}.
\bibitem[{Salda{\~n}a(2021)}]{saldana2021coding}
\bibinfo{author}{J.~Salda{\~n}a},
\newblock \bibinfo{title}{The coding manual for qualitative researchers}  (\bibinfo{year}{2021}).
\bibitem[{Treude(2024)}]{treude2024qualitative}
\bibinfo{author}{C.~Treude},
\newblock \bibinfo{title}{Qualitative data analysis in software engineering: Techniques and teaching insights},
\newblock \bibinfo{journal}{arXiv preprint arXiv:2406.08228}  (\bibinfo{year}{2024}).
\bibitem[{Tsang(2020)}]{tsang2020experiment}
\bibinfo{author}{S.~Tsang},
\newblock \bibinfo{title}{An experiment exploring the theoretical and methodological challenges in developing a semi-automated approach to analysis of small-n qualitative data},
\newblock \bibinfo{journal}{arXiv preprint arXiv:2002.04513}  (\bibinfo{year}{2020}).
\bibitem[{Achiam et~al.(2023)Achiam, Adler, Agarwal, Ahmad, Akkaya, Aleman, Almeida, Altenschmidt, Altman, Anadkat et~al.}]{achiam2023gpt}
\bibinfo{author}{J.~Achiam}, \bibinfo{author}{S.~Adler}, \bibinfo{author}{S.~Agarwal}, \bibinfo{author}{L.~Ahmad}, \bibinfo{author}{I.~Akkaya}, \bibinfo{author}{F.~L. Aleman}, \bibinfo{author}{D.~Almeida}, \bibinfo{author}{J.~Altenschmidt}, \bibinfo{author}{S.~Altman}, \bibinfo{author}{S.~Anadkat}, et~al.,
\newblock \bibinfo{title}{Gpt-4 technical report},
\newblock \bibinfo{journal}{arXiv preprint arXiv:2303.08774}  (\bibinfo{year}{2023}).
\bibitem[{Team et~al.(2023)Team, Anil, Borgeaud, Wu, Alayrac, Yu, Soricut, Schalkwyk, Dai, Hauth et~al.}]{team2023gemini}
\bibinfo{author}{G.~Team}, \bibinfo{author}{R.~Anil}, \bibinfo{author}{S.~Borgeaud}, \bibinfo{author}{Y.~Wu}, \bibinfo{author}{J.-B. Alayrac}, \bibinfo{author}{J.~Yu}, \bibinfo{author}{R.~Soricut}, \bibinfo{author}{J.~Schalkwyk}, \bibinfo{author}{A.~M. Dai}, \bibinfo{author}{A.~Hauth}, et~al.,
\newblock \bibinfo{title}{Gemini: a family of highly capable multimodal models},
\newblock \bibinfo{journal}{arXiv preprint arXiv:2312.11805}  (\bibinfo{year}{2023}).
\bibitem[{Touvron et~al.(2023)Touvron, Martin, Stone, Albert, Almahairi, Babaei, Bashlykov, Batra, Bhargava, Bhosale et~al.}]{touvron2023llama}
\bibinfo{author}{H.~Touvron}, \bibinfo{author}{L.~Martin}, \bibinfo{author}{K.~Stone}, \bibinfo{author}{P.~Albert}, \bibinfo{author}{A.~Almahairi}, \bibinfo{author}{Y.~Babaei}, \bibinfo{author}{N.~Bashlykov}, \bibinfo{author}{S.~Batra}, \bibinfo{author}{P.~Bhargava}, \bibinfo{author}{S.~Bhosale}, et~al.,
\newblock \bibinfo{title}{Llama 2: Open foundation and fine-tuned chat models},
\newblock \bibinfo{journal}{arXiv preprint arXiv:2307.09288}  (\bibinfo{year}{2023}).
\bibitem[{Sun et~al.(2023)Sun, Li, Li, Wu, Guo, Zhang, and Wang}]{sun2023text}
\bibinfo{author}{X.~Sun}, \bibinfo{author}{X.~Li}, \bibinfo{author}{J.~Li}, \bibinfo{author}{F.~Wu}, \bibinfo{author}{S.~Guo}, \bibinfo{author}{T.~Zhang}, \bibinfo{author}{G.~Wang},
\newblock \bibinfo{title}{Text classification via large language models},
\newblock in: \bibinfo{booktitle}{The 2023 Conference on Empirical Methods in Natural Language Processing}, \bibinfo{year}{2023}.
\bibitem[{Zhang et~al.(2024)Zhang, Ladhak, Durmus, Liang, McKeown, and Hashimoto}]{zhang2024benchmarking}
\bibinfo{author}{T.~Zhang}, \bibinfo{author}{F.~Ladhak}, \bibinfo{author}{E.~Durmus}, \bibinfo{author}{P.~Liang}, \bibinfo{author}{K.~McKeown}, \bibinfo{author}{T.~B. Hashimoto},
\newblock \bibinfo{title}{Benchmarking large language models for news summarization},
\newblock \bibinfo{journal}{Transactions of the Association for Computational Linguistics} \bibinfo{volume}{12} (\bibinfo{year}{2024}) \bibinfo{pages}{39--57}.
\bibitem[{Zhang et~al.(2023)Zhang, Haddow, and Birch}]{zhang2023prompting}
\bibinfo{author}{B.~Zhang}, \bibinfo{author}{B.~Haddow}, \bibinfo{author}{A.~Birch},
\newblock \bibinfo{title}{Prompting large language model for machine translation: A case study},
\newblock in: \bibinfo{booktitle}{International Conference on Machine Learning}, \bibinfo{organization}{PMLR}, \bibinfo{year}{2023}, pp. \bibinfo{pages}{41092--41110}.
\bibitem[{Brown et~al.(2020)Brown, Mann, Ryder, Subbiah, Kaplan, Dhariwal, Neelakantan, Shyam, Sastry, Askell et~al.}]{brown2020language}
\bibinfo{author}{T.~Brown}, \bibinfo{author}{B.~Mann}, \bibinfo{author}{N.~Ryder}, \bibinfo{author}{M.~Subbiah}, \bibinfo{author}{J.~D. Kaplan}, \bibinfo{author}{P.~Dhariwal}, \bibinfo{author}{A.~Neelakantan}, \bibinfo{author}{P.~Shyam}, \bibinfo{author}{G.~Sastry}, \bibinfo{author}{A.~Askell}, et~al.,
\newblock \bibinfo{title}{Language models are few-shot learners},
\newblock \bibinfo{journal}{Advances in neural information processing systems} \bibinfo{volume}{33} (\bibinfo{year}{2020}) \bibinfo{pages}{1877--1901}.
\bibitem[{Krishna et~al.(2024)Krishna, Gaur, Verma, and Jalote}]{krishna2024using}
\bibinfo{author}{M.~Krishna}, \bibinfo{author}{B.~Gaur}, \bibinfo{author}{A.~Verma}, \bibinfo{author}{P.~Jalote},
\newblock \bibinfo{title}{Using llms in software requirements specifications: An empirical evaluation},
\newblock \bibinfo{journal}{arXiv preprint arXiv:2404.17842}  (\bibinfo{year}{2024}).
\bibitem[{Nuseibeh and Easterbrook(2000)}]{nuseibeh2000requirements}
\bibinfo{author}{B.~Nuseibeh}, \bibinfo{author}{S.~Easterbrook},
\newblock \bibinfo{title}{Requirements engineering: a roadmap},
\newblock in: \bibinfo{booktitle}{Proceedings of the Conference on the Future of Software Engineering}, \bibinfo{year}{2000}, pp. \bibinfo{pages}{35--46}.
\bibitem[{Kaufmann et~al.(2022)Kaufmann, Krause, Harutyunyan, Barcomb, and Riehle}]{kaufmann2022validation}
\bibinfo{author}{A.~Kaufmann}, \bibinfo{author}{J.~Krause}, \bibinfo{author}{N.~Harutyunyan}, \bibinfo{author}{A.~Barcomb}, \bibinfo{author}{D.~Riehle},
\newblock \bibinfo{title}{A validation of qdacity-re for domain modeling using qualitative data analysis},
\newblock \bibinfo{journal}{Requirements Engineering} \bibinfo{volume}{27} (\bibinfo{year}{2022}) \bibinfo{pages}{31--51}.
\bibitem[{Chen et~al.(2016)Chen, Kocielnik, Drouhard, Pe{\~n}a-Araya, Suh, Cen, Zheng, and Aragon}]{chen2016challenges}
\bibinfo{author}{N.-C. Chen}, \bibinfo{author}{R.~Kocielnik}, \bibinfo{author}{M.~Drouhard}, \bibinfo{author}{V.~Pe{\~n}a-Araya}, \bibinfo{author}{J.~Suh}, \bibinfo{author}{K.~Cen}, \bibinfo{author}{X.~Zheng}, \bibinfo{author}{C.~R. Aragon},
\newblock \bibinfo{title}{Challenges of applying machine learning to qualitative coding},
\newblock in: \bibinfo{booktitle}{ACM SIGCHI Workshop on Human-Centered Machine Learning}, \bibinfo{year}{2016}.
\bibitem[{Glaser and Strauss(2017)}]{glaser2017discovery}
\bibinfo{author}{B.~Glaser}, \bibinfo{author}{A.~Strauss}, \bibinfo{title}{Discovery of grounded theory: Strategies for qualitative research}, \bibinfo{publisher}{Routledge}, \bibinfo{year}{2017}.
\bibitem[{Kaufmann and Riehle(2019)}]{kaufmann2019qdacity}
\bibinfo{author}{A.~Kaufmann}, \bibinfo{author}{D.~Riehle},
\newblock \bibinfo{title}{The {QDAcity-RE} method for structural domain modeling using qualitative data analysis},
\newblock \bibinfo{journal}{Requirements Engineering} \bibinfo{volume}{24} (\bibinfo{year}{2019}) \bibinfo{pages}{85--102}.
\bibitem[{Kaufmann et~al.(2020)Kaufmann, Barcomb, and Riehle}]{kaufmann:2020:supporting}
\bibinfo{author}{A.~Kaufmann}, \bibinfo{author}{A.~Barcomb}, \bibinfo{author}{D.~Riehle},
\newblock \bibinfo{title}{Supporting interview analysis with autocoding},
\newblock in: \bibinfo{booktitle}{53rd Hawaii International Conference on System Sciences, {HICSS} 2020, Maui, Hawaii, USA, January 7-10, 2020}, \bibinfo{publisher}{ScholarSpace}, \bibinfo{year}{2020}, pp. \bibinfo{pages}{1--10}.
\bibitem[{Vogelsang and Fischbach(2024)}]{vogelsang2024using}
\bibinfo{author}{A.~Vogelsang}, \bibinfo{author}{J.~Fischbach},
\newblock \bibinfo{title}{Using large language models for natural language processing tasks in requirements engineering: A systematic guideline},
\newblock \bibinfo{journal}{arXiv e-prints}  (\bibinfo{year}{2024}) \bibinfo{pages}{arXiv--2402}.
\bibitem[{Fan et~al.(2023)Fan, Gokkaya, Harman, Lyubarskiy, Sengupta, Yoo, and Zhang}]{fan2023large}
\bibinfo{author}{A.~Fan}, \bibinfo{author}{B.~Gokkaya}, \bibinfo{author}{M.~Harman}, \bibinfo{author}{M.~Lyubarskiy}, \bibinfo{author}{S.~Sengupta}, \bibinfo{author}{S.~Yoo}, \bibinfo{author}{J.~M. Zhang},
\newblock \bibinfo{title}{Large language models for software engineering: Survey and open problems},
\newblock \bibinfo{journal}{arXiv preprint arXiv:2310.03533}  (\bibinfo{year}{2023}).
\bibitem[{Bano et~al.(2024)Bano, Hoda, Zowghi, and Treude}]{bano2024large}
\bibinfo{author}{M.~Bano}, \bibinfo{author}{R.~Hoda}, \bibinfo{author}{D.~Zowghi}, \bibinfo{author}{C.~Treude},
\newblock \bibinfo{title}{Large language models for qualitative research in software engineering: exploring opportunities and challenges},
\newblock \bibinfo{journal}{Automated Software Engineering} \bibinfo{volume}{31} (\bibinfo{year}{2024}) \bibinfo{pages}{8}.
\bibitem[{Alhoshan et~al.(2023)Alhoshan, Ferrari, and Zhao}]{alhoshan2023zero}
\bibinfo{author}{W.~Alhoshan}, \bibinfo{author}{A.~Ferrari}, \bibinfo{author}{L.~Zhao},
\newblock \bibinfo{title}{Zero-shot learning for requirements classification: An exploratory study},
\newblock \bibinfo{journal}{Information and Software Technology} \bibinfo{volume}{159} (\bibinfo{year}{2023}) \bibinfo{pages}{107202}.
\bibitem[{Kici et~al.(2021)Kici, Malik, Cevik, Parikh, and Basar}]{kici2021bert}
\bibinfo{author}{D.~Kici}, \bibinfo{author}{G.~Malik}, \bibinfo{author}{M.~Cevik}, \bibinfo{author}{D.~Parikh}, \bibinfo{author}{A.~Basar},
\newblock \bibinfo{title}{A bert-based transfer learning approach to text classification on software requirements specifications.},
\newblock in: \bibinfo{booktitle}{Canadian AI}, \bibinfo{year}{2021}.
\bibitem[{Ferrari et~al.(2023)Ferrari, Spagnolo, and Gnesi}]{pure}
\bibinfo{author}{A.~Ferrari}, \bibinfo{author}{G.~O. Spagnolo}, \bibinfo{author}{S.~Gnesi},
\newblock \bibinfo{title}{Pure: a dataset of public requirements documents},
\newblock \bibinfo{journal}{Unspecified Journal} \bibinfo{volume}{Unspecified Volume} (\bibinfo{year}{2023}) \bibinfo{pages}{Unspecified Pages}.
\bibitem[{Jiang et~al.(2023)Jiang, Sablayrolles, Mensch, Bamford, Chaplot, Casas, Bressand, Lengyel, Lample, Saulnier et~al.}]{jiang2023mistral}
\bibinfo{author}{A.~Q. Jiang}, \bibinfo{author}{A.~Sablayrolles}, \bibinfo{author}{A.~Mensch}, \bibinfo{author}{C.~Bamford}, \bibinfo{author}{D.~S. Chaplot}, \bibinfo{author}{D.~d.~l. Casas}, \bibinfo{author}{F.~Bressand}, \bibinfo{author}{G.~Lengyel}, \bibinfo{author}{G.~Lample}, \bibinfo{author}{L.~Saulnier}, et~al.,
\newblock \bibinfo{title}{Mistral 7b},
\newblock \bibinfo{journal}{arXiv preprint arXiv:2310.06825}  (\bibinfo{year}{2023}).
\bibitem[{Wiegers and Beatty(2013)}]{wiegers2013software}
\bibinfo{author}{K.~E. Wiegers}, \bibinfo{author}{J.~Beatty}, \bibinfo{title}{Software requirements}, \bibinfo{publisher}{Pearson Education}, \bibinfo{year}{2013}.
\bibitem[{Coleman et~al.(2024)Coleman, Ragan, and Dari}]{coleman2024intercoder}
\bibinfo{author}{M.~L. Coleman}, \bibinfo{author}{M.~Ragan}, \bibinfo{author}{T.~Dari},
\newblock \bibinfo{title}{Intercoder reliability for use in qualitative research and evaluation},
\newblock \bibinfo{journal}{Measurement and Evaluation in Counseling and Development} \bibinfo{volume}{57} (\bibinfo{year}{2024}) \bibinfo{pages}{136--146}.
\bibitem[{Chew et~al.(2023)Chew, Bollenbacher, Wenger, Speer, and Kim}]{chew2023llm}
\bibinfo{author}{R.~Chew}, \bibinfo{author}{J.~Bollenbacher}, \bibinfo{author}{M.~Wenger}, \bibinfo{author}{J.~Speer}, \bibinfo{author}{A.~Kim},
\newblock \bibinfo{title}{Llm-assisted content analysis: Using large language models to support deductive coding},
\newblock \bibinfo{journal}{arXiv preprint arXiv:2306.14924}  (\bibinfo{year}{2023}).
\bibitem[{Turpin et~al.(2024)Turpin, Michael, Perez, and Bowman}]{turpin2024language}
\bibinfo{author}{M.~Turpin}, \bibinfo{author}{J.~Michael}, \bibinfo{author}{E.~Perez}, \bibinfo{author}{S.~Bowman},
\newblock \bibinfo{title}{Language models don't always say what they think: unfaithful explanations in chain-of-thought prompting},
\newblock \bibinfo{journal}{Advances in Neural Information Processing Systems} \bibinfo{volume}{36} (\bibinfo{year}{2024}).
\bibitem[{Wei et~al.(2022)Wei, Wang, Schuurmans, Bosma, Xia, Chi, Le, Zhou et~al.}]{wei2022chain}
\bibinfo{author}{J.~Wei}, \bibinfo{author}{X.~Wang}, \bibinfo{author}{D.~Schuurmans}, \bibinfo{author}{M.~Bosma}, \bibinfo{author}{F.~Xia}, \bibinfo{author}{E.~Chi}, \bibinfo{author}{Q.~V. Le}, \bibinfo{author}{D.~Zhou}, et~al.,
\newblock \bibinfo{title}{Chain-of-thought prompting elicits reasoning in large language models},
\newblock \bibinfo{journal}{Advances in neural information processing systems} \bibinfo{volume}{35} (\bibinfo{year}{2022}) \bibinfo{pages}{24824--24837}.

\end{thebibliography}

\end{document}